%% file: fib-paper.tex
\documentclass[12pt]{article}
\usepackage{verbatim}
\usepackage{fullpage}

\usepackage{amssymb,amsmath,amsthm}
\usepackage[mathscr]{eucal}

\usepackage{makeidx}
\input{isomath}
\input{mathenv}

\input{syms}

\usepackage{graphicx}
\usepackage{psfrag}
\usepackage{afterpage}
\usepackage{subfigure}
\usepackage{float}

\usepackage{hyperref}
\usepackage{ifpdf}

\ifpdf
\fi

\begin{document}
\title{The fixed irreducible bridge ensemble for self-avoiding walks}
\author{Michael James Gilbert}
\date{}

\maketitle
\input{abs.tex}

\newpage
\addcontentsline{toc}{section}{Contents}
\tableofcontents

\newpage
\input{body.tex}

\appendix
\newpage
\input{appx1.tex}

\bibliographystyle{acm.bst}
\bibliography{fib-paper.bib}


\end{document}

%% file: isomath.tex
\newcommand*{\bbb}[1]{\mathbb{#1}}

\newcommand{\bN}{\bbb{N}}

\newcommand{\bZ}{\bbb{Z}}

\usepackage{amsbsy}

\newcommand{\rd}{\mathrm{d}} 
\newcommand{\dif}[1]{\,\rd{#1}} 

\newcommand{\dx}{\dif{x}}



%% file: mathenv.tex
\usepackage{ifthen}

\theoremstyle{plain}
\newtheorem{thm}{Theorem}[section]

\newtheorem{con}[thm]{Conjecture}
\newtheorem*{thm*}{Theorem}
\newtheorem*{cor*}{Corollary}
\newtheorem*{lem*}{Lemma}
\newtheorem*{prop*}{Proposition}
\newtheorem*{ax*}{Axiom}
\newtheorem*{con*}{Conjecture}

\theoremstyle{definition}

\newtheorem*{defn*}{Definition}
\newtheorem*{problem*}{Problem}
\newtheorem{ex}[thm]{Example}

\newtheorem*{ex*}{Example}
\newtheorem*{exs*}{Examples}

\newtheorem*{rem*}{Remark}

\newtheorem*{aside*}{Aside}

\newtheorem*{intuition*}{Intuition}

\newtheorem*{notn*}{Notation}

\newtheorem*{conv*}{Convention}

\newtheorem*{mnem*}{Mnemonic}

\newtheorem*{motivn*}{Motivation}

\newtheorem*{exer*}{Exercise}

\numberwithin{equation}{section}

\newcommand{\beginex}{\begin{ex}$\rhd$ }
\newcommand{\mendex}{\hfill$\lhd$\end{ex}}

\newcounter{ppart}


%% file: syms.tex
\newcommand{\mcB}{\mathcal{B}}

\newcommand{\mcH}{\mathcal{H}}
\newcommand{\mcI}{\mathcal{I}}
\newcommand{\mcIn}{\mathcal{I}^n}

\newcommand{\D}{\partial}

\newcommand{\Phalf}{\mathbf{P}_{\mathbb{H},\infty}}
\newcommand{\PN}{\mathbf{P}_{\mathbb{H},N}}

\newcommand{\PSLE}{\mathbf{P}_{S,0,x+i}^{chordal}}

\newcommand{\Edelta}{\mathbf{E}_{\mathbb{H},\infty}^{\delta}}

\newcommand{\Ehalf}{\mathbf{E}_{\mathbb{H},\infty}}

\newcommand{\SLE}{\mathit{SLE}_{8/3}}

%% file: abs.tex
\begin{abstract}
We define a new ensemble for self-avoiding walks in the upper half-plane, the fixed irredicible bridge ensemble, by considering self-avoiding walks in the upper half-plane up to their $n$-th bridge height, $Y_n$, and scaling the walk by $1/Y_n$ to obtain a curve in the unit strip, and then taking $n\to\infty$. We then conjecture a relationship between this ensemble to $\SLE$ in the unit strip from $0$ to a fixed point along the upper boundary of the strip, integrated over the conjectured exit density of self-avoiding walk spanning a strip in the scaling limit. We conjecture that there exists a positive constant $\sigma$ such that $n^{-\sigma}Y_n$ converges in distribution to that of a stable random variable as $n\to\infty$. Then the conjectured relationship between the fixed irreducible bridge scaling limit and $\SLE$ can be described as follows: If one takes a SAW considered up to $Y_n$ and scales by $1/Y_n$ and then weights the walk by $Y_n$ to an appropriate power, then in the limit $n\to\infty$, one should obtain a curve from the scaling limit of the self-avoiding walk spanning the unit strip. In addition to a heuristic derivation, we provide numerical evidence to support the conjecture and give estimates for the boundary scaling exponent. 
\end{abstract}

%% file: body.tex

\section{Introduction} \label{sec:introduction}

\subsection{The infinite length upper half-plane self-avoiding walk} \label{sec:uhpsaw}

Many important 2-dimensional lattice models arise in the study of
statistical mechanics. Among these, the self-avoiding walk is
one that has been shown to be a very rich and interesting model full
that comes with a plethora of challenging problems. 

The self-avoiding walk was introduced in 1949 by Paul Flory as a model
for polymers. An $N$-step
\textit{self-avoiding walk} (SAW) on a two-dimensional lattice with
lattice spacing $\delta>0$ is a sequence of lattice sites
\begin{equation*}
\omega=\left[\omega(0),\omega(1),\ldots,\omega(N)\right]
\end{equation*} such that
$\left|\omega(j+1)-\omega(j)\right|=\delta$ for all $j=1,\ldots,N$ and
such that $\omega(j)\neq\omega(k)$ for all $j\neq k$. 
     Let $\Omega_N$ be the set of all $N$-step SAWs $\omega$ on the
lattice $\mathbb{Z}^2=\mathbb{Z}+i\mathbb{Z}$ which
begin at the origin, i.e. $\omega(0)=0$. We equip $\Omega_N$ with the
uniform probability measure, i.e. we define
$\mathbf{P}_N(\omega)=1/C_N$, where $C_N=\left|\Omega_N\right|$ is the
cardinality of $\Omega_N$. 

By concatenating an $N$-step SAW with an $M$-step SAW, we can see that 

\begin{equation}
C_{N+M}\leq C_NC_M.
\end{equation} A standard subadditivity argument then shows that there exists a constant $\mu>0$ such that 

\begin{equation} \label{connective}
\lim_{N\to\infty}\frac{\log C_N}{N}=\log\mu,
\end{equation} The constant $\mu$ is referred to as the \textit{connective constant}. 

We will be considering SAWs $\omega\in\Omega_N$ such that $\mathrm{Im}(\omega(j))>0$ for $j=1,2,\ldots$, equipped
with the uniform measure, $\mathbf{P}_{\mathbb{H},N}$. Let $\mathcal{H}_N$ denote the set of all $N$-step upper
half-plane SAWs with $\omega(0)=0$, and let $\mathcal{H}=\bigcup_{n=0}^\infty\mathcal{H}_N$ be the set of all such
upper half-plane SAWs. It was shown in \cite{lawler2002scaling} that the distributional limit as $N\to\infty$ of
the measures $\mathbf{P}_{\mathbb{H},N}$ exists and gives a measure on infinite length upper-half plane SAWs.
Let $\mathcal{H}_\infty$ denote the set of all infinite length upper half-plane SAWs, and let
$\mathbf{P}_{\mathbb{H},\infty}$ denote the weak limit of the measures $\mathbf{P}_{\mathbb{H},N}$
(considered as measures on $\mathcal{H}$. A SAW $\omega\in\mathcal{H}_N$ is called a \textit{bridge} if 

\begin{equation} \label{bridge}
\mathrm{Im}(\omega(0))<\mathrm{Im}(\omega(j))\leq\textrm{Im}(\omega(N))
\end{equation} for all $j=1,\ldots,N$. Note that the concatenation of any two bridges is still a bridge, while
not every bridge can be written as the concatenation of two shorter (non-trivial) bridges. A bridge with the
latter property will be said to be \textit{irreducible}. Let $\mathcal{B}$ denote the set of all bridges rooted
at the origin, i.e. all $\omega\in\mathcal{H}$ satisfying (\ref{bridge}) with $\omega(0)=0$, and let $\mathcal{I}$
denote the set of all irreducible bridges rooted at the origin.  

An important consequence of the proof of the existence of the limit of the measures $\mathbf{P}_{\mathbb{H},N}$
is that given an $\omega\in\mathcal{H}_\infty$, with probability 1, $\omega$ can be written as the concatenation
of a bridge with another (translated) upper half-plane SAW. The proof of the existence of
$\mathbf{P}_{\mathbb{H},\infty}$ utilizes \textbf{Kesten's relation}: 

\begin{equation} \label{Kesten}
\sum_{\omega\in\mathcal{I}}\mu^{-\left|\omega\right|}=1,
\end{equation} for the same $\mu$ as in (\ref{connective}), where $\left|\omega\right|$ denotes the length, or
number of steps, of $\omega$. Kesten's relation shows that $\mathbf{P}(\omega)=\mu^{-|\omega|}$ is a probability
measure on $\mathcal{I}$. By concatenating together an i.i.d. sequence of irreducible bridges with respect to
$\mathbf{P}$, a probability measure is induced on $\mathcal{H}$, and \cite{lawler2002scaling} shows that this
is the only possible candidate for the measure $\mathbf{P}_{\mathbb{H},\infty}$. 

One should then think of upper half-plane SAWs as having a renewal structure to them. At the end of an irreducible bridge, the future path of the walk lies entirely in the half-plane above the horizontal line where the bridge ended, and this can be considered as another infinite length upper half-plane SAW, which can then be written as the concatenation of an irreducible bridge with another infinite length half-plane SAW. 

Given any two SAWs $\omega^1$ and $\omega^2$, we will write $\omega^1\oplus\omega^2$ to denote the walk obtained by concatenating $\omega^1$ and $\omega^2$. Then the above result can be stated as follows: With $\mathbf{P}_{\mathbb{H},\infty}$-probability 1, any $\omega\in\mathcal{H}_\infty$ can be written as 

\begin{equation}
\omega=\tilde{\omega}\oplus\hat{\omega},
\end{equation} where $\tilde{\omega}\in\mathcal{I}$ and $\hat{\omega}\in\mathcal{H}_\infty$. Given a finite sequence $\omega^1,\ldots,\omega^j\in\mathcal{I}$, let $\mathcal{H}_\infty(\omega^1,\cdots,\omega^k)$ denote the set of all $\omega\in\mathcal{H}$ such that 

\begin{equation} \label{cylinder}
\omega=\omega^1\oplus\ldots\oplus\omega^k\oplus\hat{\omega},
\end{equation} where $\hat{\omega}\in\mathcal{H}_\infty$. Then we have $\mathbf{P}_{\mathbb{H},\infty}(\mathcal{H}_\infty(\omega^1,\ldots,\omega^k))=\mu^{-\sum_1^k|\omega^j|}$. For a good exposition on the bridge decomposition of self-avoiding walk, the reader is referred to \cite{dyhr2011self},\cite{madras1993self},\cite{lawler2002scaling}. 

\subsection{Summary of Results} \label{sec:results}

Consider the set of all infinite upper half-plane SAWs on the lattice $\mathbb{Z}^2=\mathbb{Z}+i\mathbb{Z}$ rooted at 0,
denoted $\mathcal{H}_\infty$, under the weak limit of the uniform measure on $\mathcal{H}_N$. Given
$\omega\in\mathcal{H}_\infty$, $\omega$ can be decomposed into the concatenation an i.i.d. sequence of irreducible
bridges $\omega^1,\omega^2,\ldots\in\mathcal{I}$. Let $Y_n=Y_n(\omega)$ denote the height of the $n$-th irreducible
bridge in the concatenation, that is
$Y_n(\omega)=\mathrm{Im}(\omega(|\omega^1\oplus\cdots\oplus\omega^n|))-\textrm{Im}(\omega(0))$ for
$\omega\in\mathcal{H}_\infty$. We conjecture that there exists $\sigma>0$ such that,

\begin{equation} \label{eq:sigma}
 \lim_{n\to\infty}\frac{Y_n}{n^\sigma} = Y,
\end{equation} where $Y$ has the distribution of a stable random variable, and the convergence here is in distribution.
That is, what equation (\ref{eq:sigma}) is really saying is that we are conjecturing that there exists $\sigma>0$ such
that $n^{-\sigma}Y_n(\omega)$ converges in distribution to that of a stable random variable as $n\to\infty$. Now given an
infinite upper half-plane SAW $\omega$, scale $\omega$ by $1/Y_n(\omega)$ to produce a curve in the unit strip and then
let $n\to\infty$. This should give a probability measure on curves in the unit strip beginning at $0$ and ending anywhere
along the upper boundary of the strip. We refer to this as the \textit{fixed irreducible bridge scaling limit}, or
\textit{fixed irreducible bridge ensemble}. It is then natural to look for some relationship between the fixed irreducible
bridge scaling limit and chordal $\mathit{SLE}_{8/3}$.   

The simplest relationship would be the following. Take an infinite upper half-plane SAW $\omega$ defined on the lattice
$\bZ^2$ and fix $n\in\bN$, some large number. Let $\hat{\omega}$ denote $\omega$ considered up to the (random) height
$Y_n(\omega)$. Scale $\hat{\omega}$ by $1/Y_n(\omega)$, so as to obtain a curve in the unit strip. In the limit
$n\to\infty$, this gives a probability measure on curves in the unit strip. Since these curves can end anywhere along
the upper boundary of the unit strip, it is necessary to integrate along the upper boundary of the strip against the
conjectured exit density for the scaling limit of SAW in the unit strip using $\mathit{SLE}$ partition functions. Let
$\rho(x)$ be the conjectured exit density for SAW in the scaling limit in the unit strip derived in \cite{dyhr2011self}
and described in Section \ref{sec:density}. Chordal $\mathit{SLE}_{8/3}$ gives a probability measure on curves in the
unit strip starting at the origin and ending at some prescribed point along the upper boundary. Thus, it might be
reasonable to ask whether the resulting measure is chordal $\mathit{SLE}_{8/3}$, integrated along the density
$\rho(x)$. In this paper, we argue that this process of scaling the walk to obtain a curve in the unit strip gives
chordal $\mathit{SLE}_{8/3}$ integrated over $\rho(x)$ if before taking the limit $n\to\infty$, we first weight the
walks by $Y_n(\omega)^p$, where the power $p$ is conjectured to be $-1/\sigma$ for $\sigma$ defined according
to (\ref{eq:sigma}), and then take the limit $n\to\infty$. The conjectured value of $\sigma$ is
$\sigma=4/3$ (see \ref{appx:sigma}).  

\subsection{Scaling limits and SLE partition functions} \label{sec:limits}

In this section we review some conjectured scaling limits of self-avoiding walk, along with \textit{SLE partition functions}, which we will use in what is to come. One, which we have already discussed, is the fixed irreducible bridge ensemble, which is obtained by considering a self-avoiding walk up to its $n$-th bridge height under the measure $\Phalf$, scaling by $1/Y_n(\omega)$ and taking $n\to\infty$. 

The next two scaling limits we consider are examples of the \textit{Schramm-Loewner evolution}, introduced by Oded Schramm in \cite{schramm2000scaling}. Let $D\subset\mathbb{C}$ be a bounded,simply connected domain (other than $\mathbb{C}$) and let $z,w\in\partial D$ be boundary points and $v\in D$ be an interior point. Given $\delta>0$, let $[z],[w],[v]$ denote the lattice points on $\delta\mathbb{Z}^2$ which are a minimum distance from $z,w$ and $v$, respectively. One can then consider all SAWs $\omega$ in $\delta\mathbb{Z}^2$ beginning at $[z]$ and ending at $[w]$, constrained to stay in $D$. We weight each walk by $\mu^{-|\omega|}$. The total weight of all such walks is then 

\begin{equation} \label{eq:total_weight}
Z_\delta(D,z;w)=\sum_{\omega\subset D:z\to w}\mu^{-|\omega|}. 
\end{equation} We then define a probability measure on all such walks $\omega$ in $D$ from $[z]$ to $[w]$ by assigning probability $\mu^{-|\omega|}/Z_\delta(D,z;w)$ to each such walk. The scaling limit as $\delta\to 0+$ is believed to exist and be equal to chordal $\SLE$ in $D$ from $z$ to $w$. We will denote the chordal $\SLE$ measure supported on curves $\gamma:[0,t_\gamma]\to\overline{D}$ such that $\gamma(0,t_\gamma)\subset D$, $\gamma(0)=z$, $\gamma(t_\gamma)=w$ by $\mathbf{P}_{D,z,w}^{chordal}$. Of particular interest to us will be the chordal $\SLE$ defined as above where $D$ is the unit strip $S:=\{z\in\mathbb{H}:0<\mathrm{Im}\ z<1\}$, $z=0$, and $w=x+i$, where $x\in\mathbb{R}$. We will denote this probability measure by $\PSLE$. 

One can also consider self-avoiding walks starting at a boundary point $[z]$ and ending at an interior point $[v]$. The resulting scaling limit is thought to be \textit{radial} $\SLE$. However, we will not be concerned with radial $\SLE$ in this paper. 

In the case that $D=\mathbb{H}$, $z=0$, and $w=\infty$, in order to obtain the scaling limit, one must first find a way to define infinite length SAWs in $\mathbb{H}$. This was done in \cite{lawler2002scaling} and is how the measure $\Phalf$ was originally defined. The scaling limit of $\Phalf$ as $\delta\to 0+$ is then conjectured to be $\mathbf{P}_{\mathbb{H},0,\infty}^{chordal}$. It is worth mentioning, however, that one can also obtain the probability measure $\Phalf$ by a method that is similar in spirit to the method for obtaining the scaling limit for SAW in bounded domains. If we consider the set of all finite length SAWs in $\mathbb{H}$ starting at $0$ and weight each such walk $\omega$ by $\mu^{-|\omega|}$, then the total weight of all such walks is infinite. If, instead, we weight each such $\omega$ by $\beta^{-|\omega|}$ for $\beta>\mu$, then the total weight is finite. The limit as $\beta\to\mu+$ has been shown to exist and to give the same measure on infinite half-plane SAWs as the weak limit on the uniform measures \cite{dyhr2011self}. 

Finally, let us consider how the normalization factor (\ref{eq:total_weight}) depends on the boundary points $z,w\in \D D$. It is conjectured that there exists a boundary scaling exponent $b>0$ and a function $H(\D D,z,w)$ such that as $\delta\to 0+$,  
\begin{equation} \label{eq:scaling_asymptotics}
Z_\delta(D,z,w)\sim \delta^{2b}H(\D D,z,w),
\end{equation} and $H(\D D,z,w)$ is thought to satisfy the following form of conformal covariance. If $\Phi$ is a conformal transformation from $D$ onto $D^{\prime}$, with $\Phi(z)=z^\prime$, $\Phi(w)=w^\prime$, then 
\begin{equation} \label{eq:conformal_covariance}
H(\D D,z,w)=|\Phi^\prime(z)|^b|\Phi^\prime(z)|^bH(\D D,z^\prime,w^\prime). 
\end{equation} \cite{lawler2002scaling,lawler2009partition,lawler16schramm}. Note that in \cite{lawler2002scaling}, the boundary scaling exponent is denoted by $a$, whereas we are denoting it by $b$. 

Recently, it has been shown that there are lattice effects which should persist in the scaling limit for general domains $D$ \cite{kennedy2011lattice}. Therefore, one cannot expect equations (\ref{eq:scaling_asymptotics}) and (\ref{eq:conformal_covariance}) to provide a full description of the scaling limit for general domains $D\subset\mathbb{C}$. However, we will be restricting our attention to curves in the domains $\mathbb{H}$ and $S$, for which there are no lattice effects expected to persist in the scaling limit. 

In section \ref{sec:predictions} we will use equation (\ref{eq:conformal_covariance}) to derive the predicted exit density for the scaling limit of self-avoiding walks in the unit strip beginning at the origin and ending anywhere along the upper boundary. We will denote the density by $\rho(x)$, where we are assuming that each walk exits the strip at some point $x+i$ with $x\in\mathbb{R}$. 

In section \ref{sec:conjectures} we state our conjecture about how to obtain chordal $\SLE$ from the fixed irreducible bridge ensemble precisely and provide a heuristic argument. The conjecture involves the stability parameter, $\sigma$, defined according to (\ref{eq:sigma}). In order to test this conjecture (section \ref{sec:simulations}), we require a definite value for $\sigma$. We conjecture that $\sigma=4/3$. In the Appendix \ref{appx:sigma}, we present a heuristic argument, originally due to Tom Kennedy via private communication, in support of this.   

\section{The conjecture} \label{sec:conjectures}

\subsection{Statement of the conjecture} \label{sec:statement}
In order to precisely state our conjecture, we first recall some notations introduced in section \ref{sec:introduction}. $\PN$ denotes the probability measure on $N$-step upper-half plane SAWs beginning at $0$ defined on the lattice $\mathbb{Z}^2$, and $\Phalf$ denotes the probability measure on infinite length SAWs in the upper half plane beginning at $0$ and ending at $\infty$, defined on the lattice $\mathbb{Z}^2$. $\PSLE$ denotes chordal $\SLE$ measure in the unit strip $S$ on curves beginning at $0$ and ending at $x+i$, and $\rho(x)$ denotes the exit density along the upper boundary $\mathrm{Im}(z)=1$ of the scaling limit for SAW in the unit strip $S$, starting at $0$ and ending anywhere along the upper boundary. Then the conjecture can be stated as follows:
\begin{con} \label{th:main} The fixed irreducible bridge scaling limit of the SAW and chordal $\SLE$ in the unit strip $S$ are related by 
\begin{equation}
\lim_{n\to\infty}\frac{\Ehalf\left[Y_n(\omega)^{-1/\sigma}1\left(\hat{\omega}/Y_n(\omega)\in E\right)\right]}{\Ehalf\left[Y_n(\omega)^{-1/\sigma}\right]} = \int_{-\infty}^\infty\dx \rho(x)\PSLE(E),
\end{equation} where $E$ is an event of simple curves in the unit strip $S$ beginning at $0$ and ending anywhere along the upper boundary of the strip, $\omega$ is an infinite upper half-plane SAW, and $\hat{\omega}$ is the curve $\omega$ considered up to the time it reaches height $Y_n(\omega)$, the $n$th bridge height of $\omega$. 
\end{con} So we can generate chordal $\SLE$ in the unit strip by generating an $N$-step SAW $\omega$ for very large values of $N$, considered up to height $Y_n(\omega)$ for large values of $n$, scaled by $1/Y_n(\omega)$, and then giving it the weight $Y_n(\omega)^{-1/\sigma}$. The conjectured value of $\sigma$ is 4/3. 
\subsection{The derivation} \label{sec:derivation}
In order to derive Conjecture \ref{th:main}, we fix two heights $y_1$ and $y_2$, which we think of as order 1, and a large real number $L>0$. We will then only consider curves which have a bridge point in the region $A=\{z\in\mathbb{H}:y_1L\leq\mathrm{Im}(z)\leq y_2L\}$. Let $\mcIn = \mcI\times\cdots\times\mcI$ ($n$ times) be the set of all $\omega\in\mcH_\infty$ such that $\omega=\omega^1\oplus\cdots\oplus\omega^n$, with $\omega^1,\ldots,\omega^n\in\mcI$, i.e. the set of all concatenations of $n$ irreducible bridges beginning at the origin. Recall that if $\hat{\omega}\in\mcIn$ and $\mcH_\infty(\hat{\omega})$ denotes the set of all $\omega\in\mcH_\infty$ such that $\omega=\hat{\omega}\oplus\tilde{\omega}$ with $\tilde{\omega}\in\mcH_\infty$, then we have 
\begin{equation*}
\Phalf\left(\mcH_\infty(\hat{\omega})\right) = \mu^{-|\hat{\omega}|}. 
\end{equation*} Therefore, the total weight of all SAWs in $\mcH_\infty$ with a bridge pont in $A$ is 
\begin{equation} \label{eq:Zdef}
Z(A)=\sum_{n=0}^\infty\sum_{\hat{\omega}\in\mcIn}\mu^{-|\hat{\omega}|}1\left(Y_n(\omega)\in[y_1L,y_2L]\right).
\end{equation}  Now let $E$ be an event of simple curves in the strip $S$ starting at $0$ and ending anywhere along the upper boundary of the strip. We define the probability of the event $E$ to be $N(E,A)/Z(A)$, where 
\begin{equation} \label{eq:Ndef} 
N(E,A)=\sum_{n=0}^\infty\sum_{\hat{\omega}\in\mcIn}\mu^{-|\hat{\omega}|}1\left(Y_n(\omega)\in[y_1L,y_2L]\right)1\left(\hat{\omega}/Y_n(\omega)\in E\right). 
\end{equation} According to the definition of $\Phalf$, we have 
\begin{equation} \label{eq:step1}
N(E,A)=\sum_{n=0}^\infty\Phalf\left[1\left(Y_n(\omega)\in[y_1L,y_2L]\right)1\left(\hat{\omega}/Y_n(\omega)\in E\right)\right]. 
\end{equation} Since we have fixed $L$ to be a very large number, this forces each term in the above sum to be zero other than those corresponding to very large values of $n$. Then, according to \ref{eq:sigma}, if we fix $N\in\bN$ large enough, $N^{-\sigma}Y_N$ should have approximately the same distribution as $n^{-\sigma}Y_n$ for all $n$ sufficiently large. Therefore, the condition $Y_n(\omega)\in[y_1L,y_2L]$ can be replaced with the condition (for very large fixed $N$)
\begin{equation} \label{eq:new_condition}
y_1Ln^{-\sigma}N^{\sigma}\leq Y_N(\omega)\leq y_2Ln^{-\sigma}N^{\sigma}.
\end{equation} Furthermore, since for large values of $n$, the distribution of $\hat{\omega}/Y_n(\omega)$ approaches the distribution of a curve pulled from the fixed irreducible bridge ensemble, the condition $\hat{\omega}/Y_n(\omega)\in E$ can be replaced with the condition $\hat{\omega}/Y_N(\omega)\in E$. This, along with (\ref{eq:step1}) and (\ref{eq:new_condition}) lead to 
\begin{equation}
N(E,A)\approx\sum_{n=0}^\infty\Ehalf\left[1\left(\left(N^\sigma\frac{y_1L}{Y_N(\omega)}\right)^{1/\sigma}\leq n\leq\left(N^\sigma\frac{y_2L}{Y_N(\omega)}\right)^{1/\sigma}\right)1\left(\hat{\omega}/Y_N(\omega)\in E\right)\right]. 
\end{equation} Now we move the sum on $n$ inside the expectation and consider 
\begin{equation*}
\sum_{n=0}^\infty1\left(\left(N^\sigma\frac{y_1L}{Y_N(\omega)}\right)^{1/\sigma}\leq n\leq\left(N^\sigma\frac{y_2L}{Y_N(\omega)}\right)^{1/\sigma}\right). 
\end{equation*} This sum is easy to approximate. We have
\begin{align*}
\sum_{n=0}^\infty\mathbf{1}\left(\left(N^\sigma\frac{y_1L}{Y_n(\omega)}\right)^{-1/\sigma}\leq n\leq\left(N^\sigma\frac{y_2L}{Y_n(\omega)}\right)^{-1/\sigma}\right) & \approx L^{1/\sigma}Y_N(\omega)^{-1/\sigma}N(y_2-y_1) \\
 & = cNL^{1/\sigma}Y_n(\omega)^{-1/\sigma},
\end{align*} where $c=y_2-y_1$. The factor of $cNL^{1/\sigma}$ will cancel out of both numerator and denominator, and what we are left with is 
\begin{equation} \label{eq:fib_probability}
\lim_{N\to\infty}\frac{N(E,A)}{Z(A)}=\lim_{n\to\infty}\frac{\Ehalf\left[Y_n(\omega)^{-1/\sigma}1\left(\hat{\omega}/Y_n(\omega)\in E\right)\right]}{\Ehalf\left[Y_n(\omega)^{-1/\sigma}\right]}. 
\end{equation}
Next, we decompose the sum by the value of the bridge heights. Given a SAW $\omega\in\mcH_\infty$, let $D=D(\omega)$ be the set of bridge heights. That is, $D(\omega)$ is the set of all $y\geq 0$ such that there exists $n=0,1,\ldots$ such that $Y_n(\omega)=y$. Then we have 
\begin{align*}
N(E,A) & =\sum_{n=0}^\infty\sum_{\hat{\omega}\in\mcIn}\mu^{-|\hat{\omega}|}1\left(\hat{\omega}/Y_n(\omega)\in E\right)1\left(Y_n(\omega)\in[y_1L,y_2L]\right) \\
& = \sum_{y\in\mathbb{Z}\cap[y_1L,y_2L]}\sum_{n=0}^\infty\sum_{\hat{\omega}\in\mcIn}\mu^{-|\hat{\omega}|}1(\hat{\omega}/y\in E)1(Y_n=y) \\
& = \sum_{y\in\mathbb{Z}\cap[y_1L,y_2L]}\Phalf(\hat{\omega}/y\in E;y\in D) \\
& = \sum_{y\in\mathbb{Z}\cap[y_1L,y_2L]}\Phalf(\hat{\omega}/y\in E|y\in D)\Phalf(y\in D). 
\end{align*} Similarly, we find that 
\begin{equation*}
Z(A) = \sum_{y\in\mathbb{Z}\cap[y_1L,y_2L]}\Phalf(y\in D). 
\end{equation*} In \cite{dyhr2011self}, it was shown that conditioning on the event that a SAW $\omega\in\mcH_\infty$ has a bridge height at $y$ and considering the walk up to height $y$ gives the law for self-avoiding walk in the strip $\{z\in\mathbb{H}:0<\mathrm{Im}\ z<y\}$. Therefore, by taking $y$ large enough, and scaling the walk by $1/y$, one should expect to get a distribution approaching that of the distribution of SAW in the unit strip $S$ starting at $0$ and ending anywhere along the upper boundary of the strip, in the scaling limit. It follows that if we sum $\Phalf(\hat{\omega}/y\in E|y\in D)$ over all $y\in\bZ\cap [y_1L,y_2L]$, then take the limit $L\to\infty$, which is effectively the same as taking the limit $N\to\infty$ in \ref{eq:fib_probability}, $N(E,A)/Z(A)$ should converge to the law for $\SLE$ in the unit strip, starting at $0$ and ending at $x+i$, $x\in\mathbb{R}$, integrated over the density $\rho(x)$. In other words, we have  
\begin{align} \label{eq:SLE_probability}
\lim_{L\to\infty}\frac{N(E,A)}{Z(A)} & = \lim_{L\to\infty}\frac{\sum_{y\in\mathbb{Z}\cap[y_1L,y_2L]}\Phalf(\hat{\omega}/y\in E|y\in D)\Phalf(y\in D)}{\sum_{y\in\mathbb{Z}\cap[y_1L,y_2L]}\Phalf(y\in D)} \\
& \approx \lim_{L\to\infty}\frac{c\Phalf(\hat{\omega}/L\in E|L\in D)\Phalf(L\in D)}{c\Phalf(L\in D)} \\ 
& = \int_{-\infty}^\infty\dx\rho(x)\PSLE(E). 
\end{align} This completes the derivation of conjecture \ref{th:main}. 

\section{SLE predictions of random variables} \label{sec:predictions}

\subsection{The density function $\rho(x)$} \label{sec:density}
In this section we will use $\mathit{SLE}$ partition functions to derive a conjecture for the exit density, $\rho(x)$, for the scaling limit of self-avoiding walk defined on the unit strip, along the upper boundary. Recall that for a simply connected domain $D$ and points $z,w\in\D D$, the $\mathit{SLE}$ partition function $H(D,z,w)$ satisfies the conformal covariance property (\ref{eq:conformal_covariance}). It was predicted in \cite{lawler2002scaling} that the boundary scaling exponent for SAW is $b=5/8$. Using this value, the conformal covariance property takes the following form: If $\Phi$ is any conformal transformation, then 
\begin{equation} \label{conf_cov2}
H(D,z,w)=\left|\Phi^\prime(z)\Phi^\prime(w)\right|^{5/8}H(\Phi(D),\Phi(z),\Phi(w)). 
\end{equation} This defines $H(D,z,w)$ up to specifying it for a particular choice of domain $D$ and boundary points $z$ and $w$. The convention we follow is of taking $H(\mathbb{H},0,1)=1$. 

First note that if we take $\Phi$ to be a dilation $\Phi(z)=xz$, for $x\in\mathbb{R}\setminus\{0\}$, then by (\ref{conf_cov2}), we have 
\begin{equation} \label{eq:halfplane_x}
H(\mathbb{H},0,x)=\left(\frac{1}{x^2}\right)^{5/8}=\frac{1}{x^{5/4}}. 
\end{equation} Therefore we can calculate $H(S,0,x+i)$ by considering the conformal map $f:S\to\mathbb{H}$ such that $f(0)=0$, $f(x+i)=-e^{\pi x}-1$, given by $f(z)=e^{\pi z}-1$. We have $\left|f^\prime(0)\right|=\pi$ and $\left|f^\prime(x+i)\right|=\pi e^{\pi x}$. Thus, 
\begin{align*} 
H(S,0,x+i) & = \left|f^\prime(0)\right|^{5/8}\left|f^\prime(x+i)\right|^{5/8}H(\mathbb{H},0,-e^{\pi x}-1) \\
 & = \left[\frac{\pi^2e^{\pi x}}{\left(1+e^{\pi x}\right)^2}\right]^{5/8} = \left[\frac{\pi^2}{\cosh^2(\pi x/2)}\right]^{5/8}. 
\end{align*} According to (\ref{eq:scaling_asymptotics}), this shows that the probability density function $\rho(x)$ should be given by 
\begin{equation} \label{eq:rho_cform}
\rho(x) = c\left[\cosh\left(\frac{\pi x}{2}\right)\right]^{-5/4}, 
\end{equation} where $c$ is a normalization constant. 
\subsection{The right-most excursion} \label{sec:rightmost}

Given a SAW $\omega$ defined in the unit strip $S$ with lattice spacing $\delta$, let $X(\omega)=\max_{j}\mathrm{Re}\ \omega(j)$ denote the \textit{rightmost excursion} of $\omega$, i.e. the right-most point on the SAW in the strip. Based on the results of Section \ref{sec:density}, we conjecture that, in the scaling limit, $X$ has distribution given by 
\begin{equation} \label{eq:rightmost_distribution}
\lim_{\delta\to0+}\mathbf{P}\left(X<\xi\right) = \int_{-\infty}^\infty\PSLE\left(\max_t\mathrm{Re}\ \gamma(t)<\xi\right)\rho(x)\dx, 
\end{equation} where $\gamma(t)$ is an $\SLE$ curve in the unit strip, starting at $0$ and ending at $x+i$, and $\rho(x)$ is given by (\ref{eq:rho_cform}). To calculate $\PSLE\left(\max_t\mathrm{Re}(\gamma(t))<\xi\right)$, we use the following form of \textit{conformal invariance}: If $D$ is a simply connected domain, $z,w\in\D D$, and $f:D\to D^\prime$ is a conformal transformation, 
\begin{equation} \label{eq:conformal_invarianceSLE}
\mathbf{P}^{chordal}_{D,z,w}\left(\gamma[0,\infty)\cap A =\emptyset\right) = \mathbf{P}^{chordal}_{D^\prime,f(z),f(w)}\left(\tilde{\gamma}[0,\infty)\cap f(A) = \emptyset\right),
\end{equation} where $\mathbf{P}^{chordal}_{D,z,w}$ denotes chordal $\SLE$ measure in $D$, starting at $z$ and ending at $w$, $\mathbf{P}_{D^\prime,f(z),f(w)}^{chordal}$ denotes chordal $\SLE$ measure in $D^\prime$, starting at $f(z)$ and ending at $f(w)$, and $A$ is a closed set such that $A\subset \overline{D}$, $z,w\notin A$, $A\cap \overline{D}\subset \partial D$ and $D\setminus A$ is simply connected. 

Conformal invariance is built into the definition of every $\mathit{SLE}_\kappa$ measure. However, $\SLE$ measure also satisfies the following restriction property: If $D$ is a simply connected domain, $z,w\in\partial D$, and $D^\prime\subset D$ is another simply connected domain with $z,w\in\partial D^\prime$, then 
\begin{equation} \label{eq:conformal_restrictionSLE}
\mathbf{P}_{D,z,w}^{chordal}\left(\gamma[0,\infty)\cap A = \emptyset|\gamma[0,\infty)\subset D^\prime\right) = \mathbf{P}_{D^\prime,z,w}^{chordal}\left(\tilde{\gamma}[0,\infty)\cap A = \emptyset\right),
\end{equation} where $A$ is as in (\ref{eq:conformal_invarianceSLE}), along with the assumption that $A\subset \overline{D^\prime}$. In \cite{lawler2003conformal}, \cite{lawler2008conformally}, it is shown that for any probability measure on a certain type of random subsets in the plain called \textit{restriction hulls}, which satisfy (\ref{eq:conformal_invarianceSLE}), (\ref{eq:conformal_restrictionSLE}), the probability in (\ref{eq:conformal_invarianceSLE}) can be computed by 
\begin{equation} 
\mathbf{P}_{\mathbb{H},0,\infty}^{chordal}\left(K\cap A=\emptyset\right) = \Phi_A^\prime(0)^\alpha, \label{eq:Virag_formula}
\end{equation} for some real number $\alpha$, where $K$ is a restriction hull and $\Phi_A$ is the unique conformal transformation mapping $\mathbb{H}\setminus A$ onto $\mathbb{H}$ with $\Phi_A(0)=0$, $\Phi_A(\infty) = \infty$, and $\Phi_A(z)=z+ o(1)$ as $z\to\infty$. 

In the case of $\SLE$, it is known that $\alpha = 5/8$, and therefore we have 
\begin{equation}
\mathbf{P}_{\mathbb{H},0,\infty}^{chordal}\left(\gamma[0,\infty)\cap A=\emptyset\right) = \Phi_A^\prime(0)^{5/8}. \label{eq:SLEprob}
\end{equation}

It is well known that the map $f(z)=e^{\pi x}$ defines a conformal transformation from the unit strip to the half-plane
$\mathbb{H}$ satisfying $f(0)=1$, $f(x+i)= -e^{\pi x}$. Therefore, the map 
\begin{equation}
\Psi_x(z) = \frac{e^{\pi z}-1}{e^{\pi z} + e^{\pi x}} \label{eq:psidef}
\end{equation} defines a conformal transformation from $S$ onto $\mathbb{H}$ with $\Psi_x(0) = 0$ and $\Psi_x(x+i) = \infty$.
It follows then from (\ref{eq:conformal_invarianceSLE}) that if $x<\xi$, 
\begin{align*}
\PSLE\left(\max_t\textrm{Re}(\gamma(t))<\xi\right) & = \PSLE\left(\gamma[0,\infty)\cap\{z\in S:\textrm{Re}(z)\geq\xi\}=\emptyset\right) \\
& = \mathbf{P}_{\mathbb{H},0,\infty}^{chordal}\left(\tilde{\gamma}[0,\infty)\cap \Psi_x\left(\{z\in S:\textrm{Re}(z)\geq\xi\}\right)=\emptyset\right). 
\end{align*}

Let $A = \Psi_x\left(\{z\in S:\textrm{Re}(z)\geq\xi\}\right)$. Then we can write $A = \{z\in\mathbb{H}:|z-c(x,\xi)|\leq a(x,\xi)\}$, where 
\begin{align*}
c(x,\xi) & = \frac{1}{2}\left(\frac{e^{\pi\xi}+1}{e^{\pi\xi}-e^{\pi x}}+\frac{e^{\pi\xi}-1}{e^{\pi\xi}+e^{\pi x}}\right) \\ 
a(x,\xi) & = \frac{1}{2}\left(\frac{e^{\pi\xi}+1}{e^{\pi\xi}-e^{\pi x}}-\frac{e^{\pi\xi}-1}{e^{\pi\xi}+e^{\pi x}}\right). 
\end{align*} 

In this case we can write down $\Phi_A$ explicitly. We have 
\begin{equation} 
\Phi_A(z) = (z-c(x,\xi))+\frac{a(x,\xi)^2}{z-c(x,\xi)}. \label{eq:PhiA}
\end{equation} Evaluating the derivative of (\ref{eq:PhiA}) at $0$ and using (\ref{eq:SLEprob}), we find that 
\begin{align*}
\PSLE\left(\max_t\textrm{Re}(\gamma(t))<\xi\right) & = \Phi_A^\prime(0)^{5/8} \\
& = \left[1-\left(\frac{a(x,\xi)}{c(x,\xi)}\right)^2\right]^{5/8}.
\end{align*} 

Therefore, we can calculate the distribution of the right most excursion of SAW in the strip in the scaling limit by 
\begin{equation*}
\lim_{\delta\to0+}\mathbf{P}\left(X<\xi\right) = \int_{-\infty}^\xi\left[1-\left(\frac{a(x,\xi)}{c(x,\xi)}\right)^2\right]^{5/8}\rho(x)\dx, 
\end{equation*} where $\rho(x)$ is given by (\ref{eq:rho_cform}). Thus, by our Conjecture, \ref{th:main}, we should have 
\begin{equation}
\lim_{n\to\infty}\frac{\Edelta\left[Y_n(\omega)^{-1/\sigma}1(\max_j\textrm{Re}(\omega(j))/Y_n(\omega)<\xi)\right]}{\Edelta\left[Y_n(\omega)^{-1/\sigma}\right]} = \int_{-\infty}^\xi\left[1-\left(\frac{a(x,\xi)}{c(x,\xi)}\right)^2\right]^{5/8}\rho(x)\dx. \label{eq:rightmost_conjecture}
\end{equation}

\section{Simulations} \label{sec:simulations}

The pivot algorithm provides us with a fast chain Monte Carlo algorithm for simulating the self-avoiding
walk in the full plane or the half-plane. It has also recently been shown in \cite{dyhr2011self} that the
pivot algorithm can be used to simulate self-avoiding walks in the strip $S$. Taking lattice effects into
account (see \cite{kennedy2011lattice}), it should also be possible to simulate the self-avoiding walk in
other domains using the pivot algorithm. Recently, Nathan Clisby has developed a very fast implementation
of the pivot algorithm, \cite{clisby2010efficient}, and that is the algorithm that we use for our simulations. 

We use the pivot algorithm to generate self-avoiding walks in the half-plane with number of steps $N=1$ million.
Each iteration of the algorithm is highly correlated, so there is no point in sampling each iteration. Instead,
we sample every 100 iterations. In this way, we generated $144$ million samples. 

We first test the conjectured density $\rho(x)$ given by (\ref{eq:rho_cform}) against the sampled data. We take
$n=100$, sample self-avoiding walks in the half-plane, considering them up to their 100th bridge point, and then
scale them by $1/Y_n$ to get a curve in the unit strip. To test the exit density of these curves against $\rho(x)$,
we split the interval $[-3,3]$ into 600 equal parts of length $\dx = 0.01$. We then plot a histogram by summing the
weights $Y_n^{-1/\sigma}$ for each curve $\omega$ sampled which satisfies $x\leq\textrm{Re}(\omega(s)/Y_n)<x+\dx$,
divided by the sum of the weights $Y_n^{-1/\sigma}$ for every curve sampled. Here we are using $s$ to denote the time
at which $\omega$ reaches height $Y_n$. We have also plotted a histogram of the exit density of the curves in the strip
obtained from our samples by normalizing by the number of samples generated instead of the sum of the weights $Y_n^{-1/\sigma}$
in order to show that we do not get the conjectured exit density $\rho$. 

\begin{figure}[!htb]
\begin{center}
\includegraphics[height=8cm]{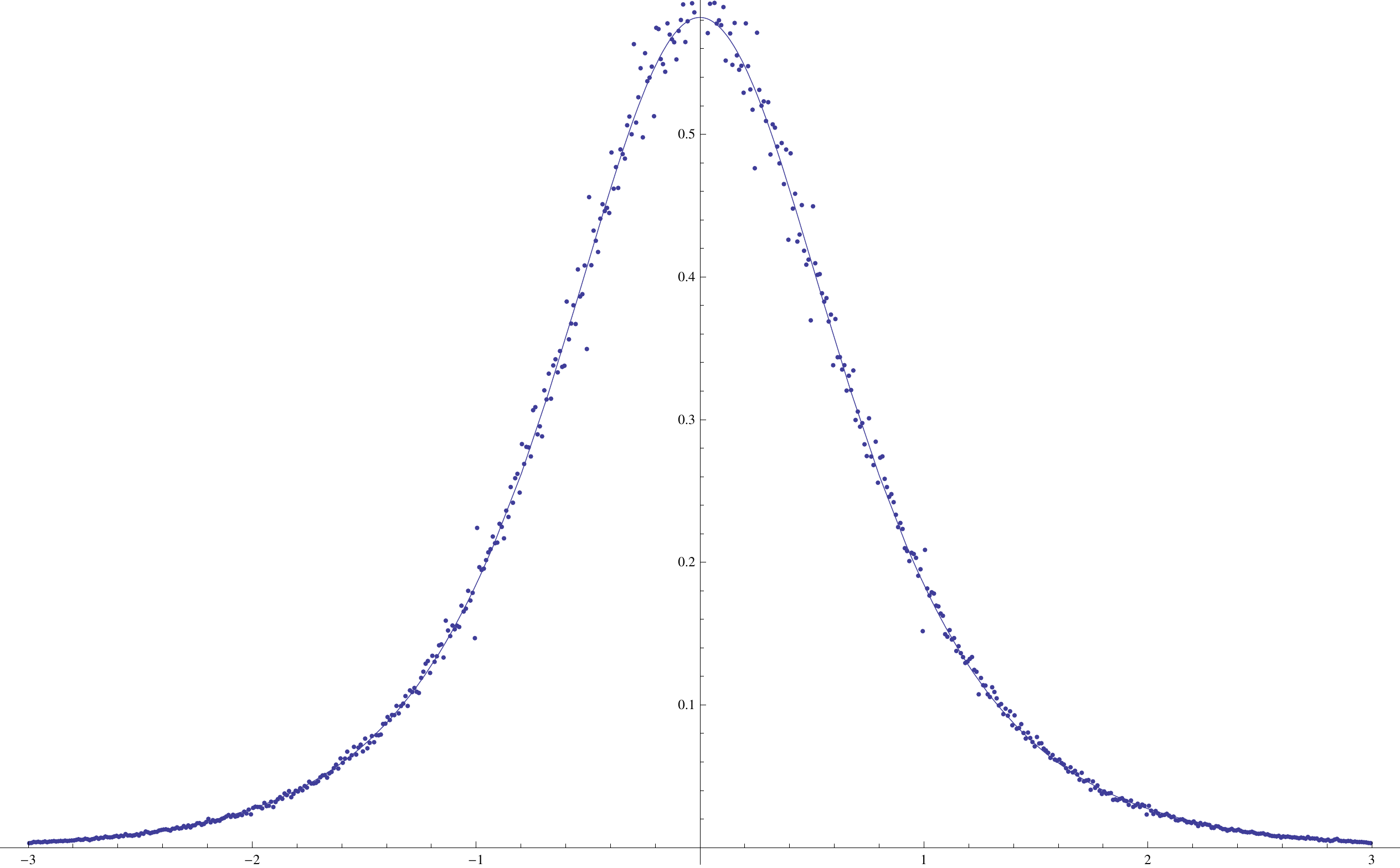}
\caption{Histogram of exit points along the upper boundary of the strip for the fixed irreducible bridge ensemble.
The conjectured density $\rho(x)$ is represented by the solid curve, while the histogram is represented by the data
points. \label{fig:good_density}}
\end{center}
\end{figure}

\begin{figure}[!htb]
\begin{center}
\includegraphics[height=8cm]{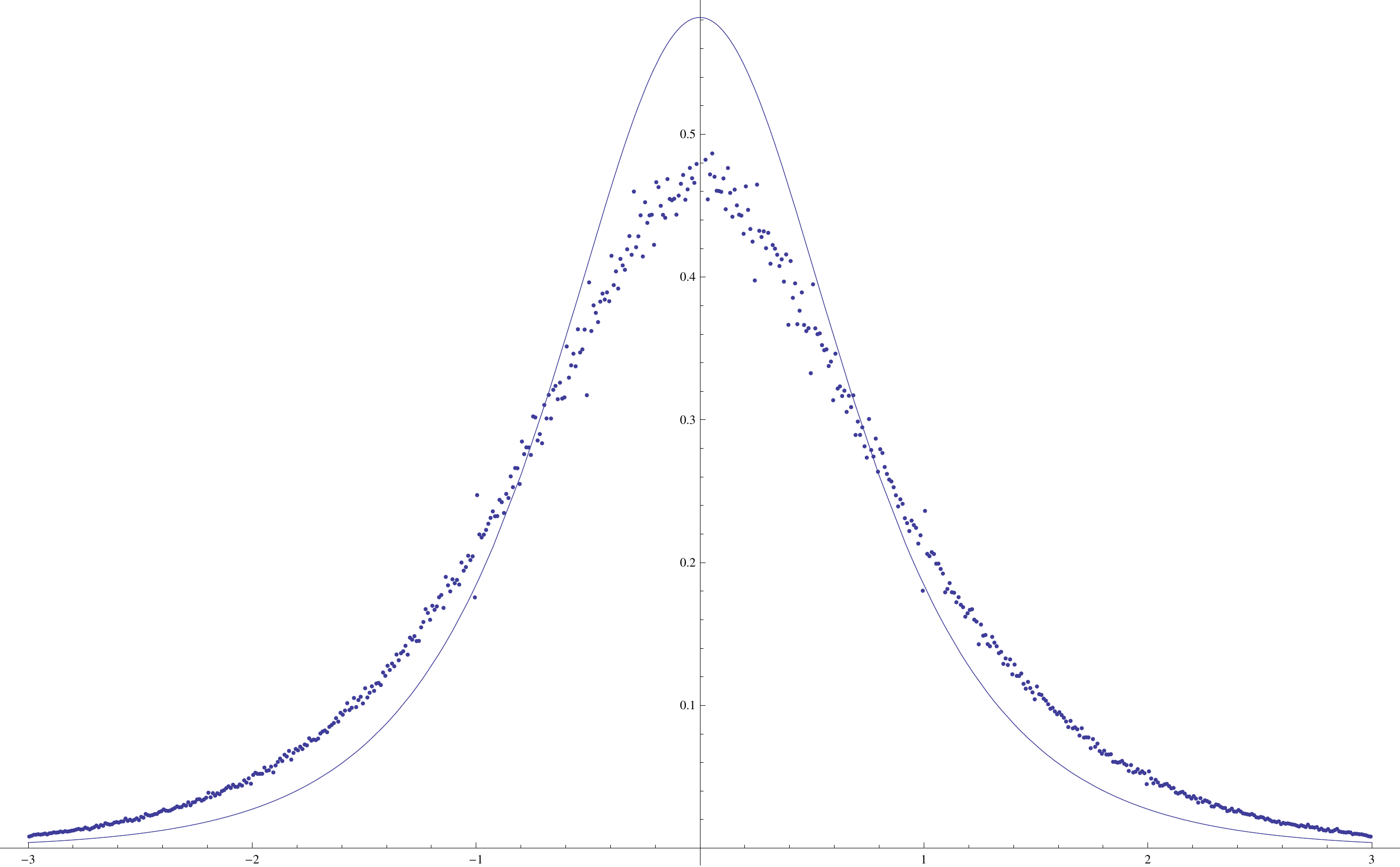}
\caption{Histogram of exit points along the upper boundary of the strip obtained by normalizing by the number of samples generated,
as opposed to normalizing by the sum of the weights $Y_n^{-1/\sigma}$. Once again $\rho(x)$ is represented by the solid curve.
\label{fig:bad_density}}
\end{center}
\end{figure}

Next we test the conjecture by making a prediction for the scaling exponent $b$ in (\ref{eq:scaling_asymptotics}) and
({\ref{eq:conformal_covariance}). We do this by plotting the $\log$ of
$\mathbf{E}_N[Y_n^{-1/\sigma}1(x\leq\textrm{Re}(\omega(s)/Y_n)<x+\dx)]$
versus the $\log$ of $\cosh^{-2}(\pi(x+\dx/2)/2)$.
We take evenly spaced values of the interval $[-1.90,1.90]$ with spacing $\dx = 0.01$. By Conjecture \ref{th:main}, we should have 
\begin{equation}
\log\left(\mathbf{E}_N[Y_n^{-1/\sigma}1(x\leq\textrm{Re}(\omega(s)/Y_n)<x+\dx)]\right) = b\log\left(\cosh^{-2}(\pi(x+\dx/2)/2)\right)+\textrm{const}. \label{eq:logeq}
\end{equation} Therefore, the data points should lie on a line. The slope of the line should be $b$,
which is conjectured to be $5/8 = 0.625$. 

\begin{figure}[!htb]
\begin{center}
\includegraphics[height=8cm]{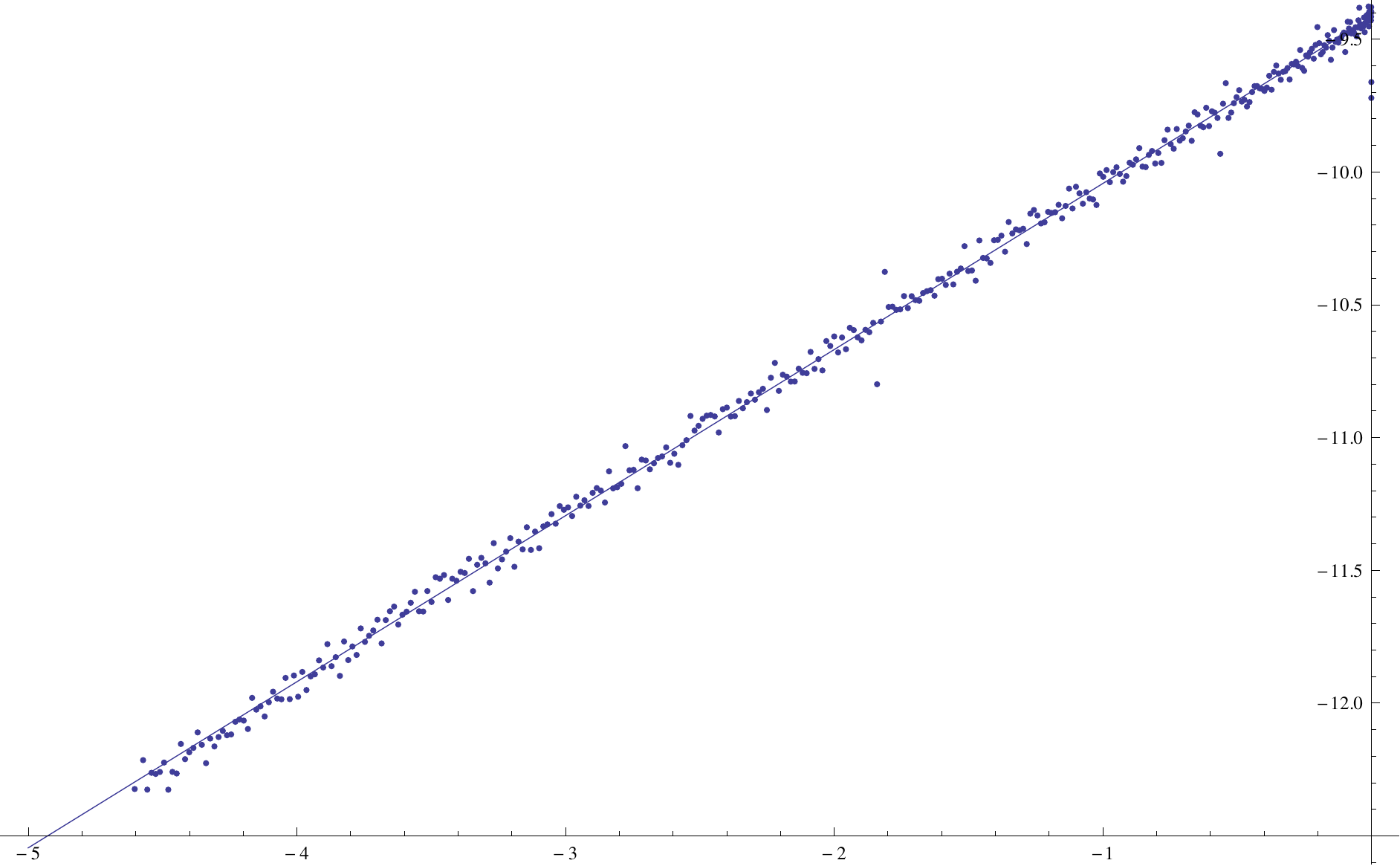}
\caption{log-log plot of $\mathbf{E}_N[Y_n^{-1/\sigma}1(x\leq\textrm{Re}(\omega(s)/Y_n)<x+\dx)]$ versus $\cosh^{-2}(\pi(x+\dx/2)/2)$.
The slope of the least-squares fit is 0.625303.\label{fig:good_logplot}}
\end{center}
\end{figure}

The line shown in Figure \ref{fig:good_logplot} was calculated using unweighted least-squares. No attempt was made
to find an error estimate. The slope of the least-squares fit
is 0.625303. If we let $b$ denote the slope of our least-squares fit, then comparing $b$ with the conjectured value, we have 
\begin{equation}
b-5/8 = 0.000303. 
\end{equation} 
We have also plotted a log-log graph of the expected value of $1(x\leq\textrm{Re}(\omega(s)/Y_n)<x+\dx)$ versus $\cosh^{-2}(\pi(x+\dx/2)/2$ by calculating the expected value through the number of samples as opposed to summing the weights $Y_n^{-1/\sigma}$. This should be compared to Figure \ref{fig:good_logplot}. The slope of the least-squares fit in this case is 0.444367. 

\begin{figure}[!htb]
\begin{center}
\includegraphics[height=8cm]{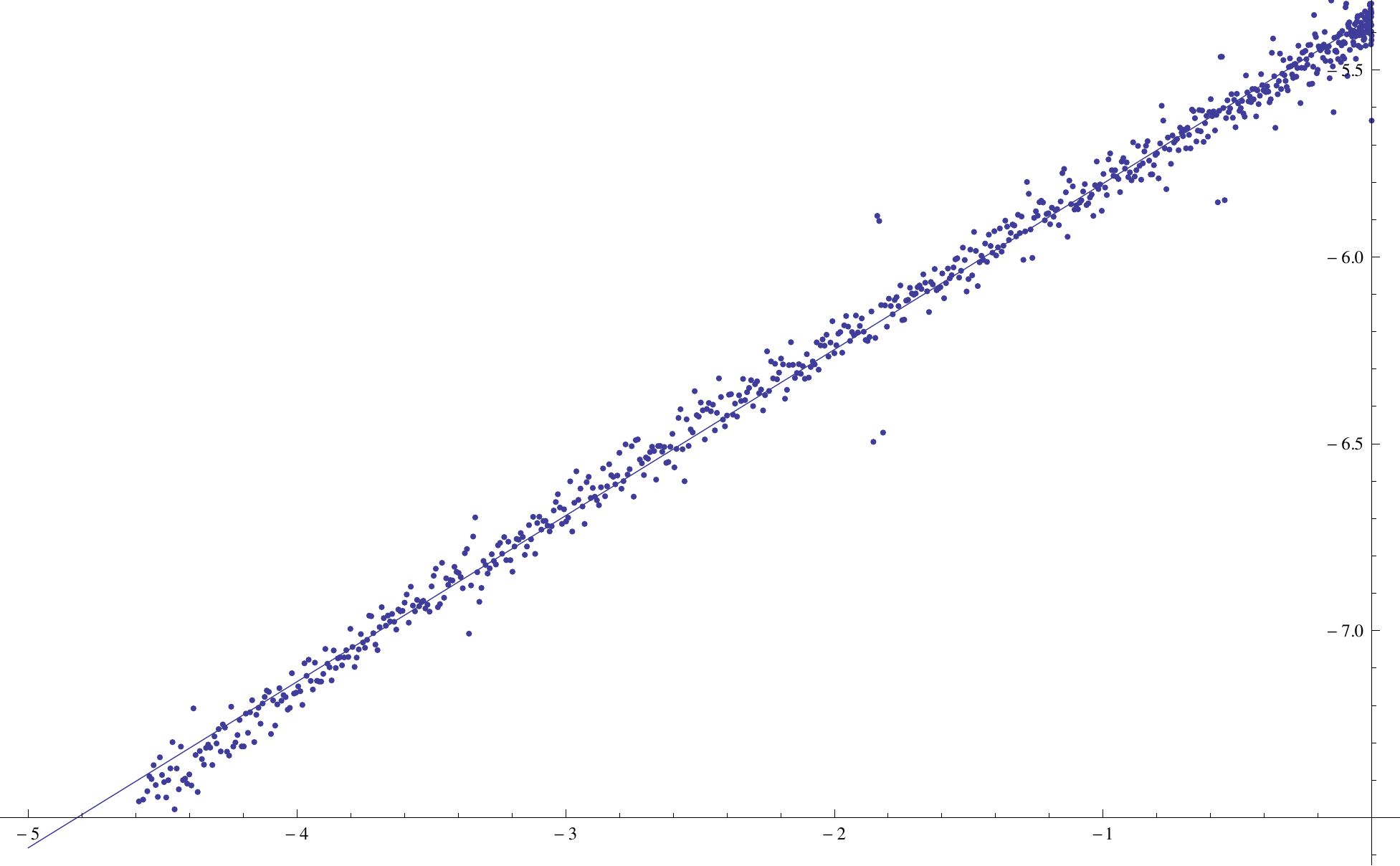}
\caption{log-log plot of $\mathbf{E}_N[1(x\leq\textrm{Re}(\omega(s)/Y_n)<x+\dx)]$ versus $\cosh^{-2}(\pi(x+\dx/2)/2)$. Without taking the weights $Y_n^{-1/\sigma}$ into account, the slope of the least-squares fit is 0.444367. \label{fig:bad_logplot}}
\end{center}
\end{figure}

Next, we perform numerical tests on the rightmost excursion, which we are denoting by $X$. After generating our
self-avoiding walks in the half-plane with $N=1$ million steps, and considering them up to their 100th bridge point (i.e. taking $n=100$),
we scale the walks by $1/Y_n$ and weight the probability measure by $Y_n^{-1/\sigma}$. We denote the probability obtained in this manner by $\mathbf{P}_{N,n}$. Of course this depends on the number of steps in the walk, as well as the value of $n$. But for large enough values of $N,n$, this measure should look very close to the fixed irreducible bridge measure. Conjecture \ref{th:main} then states that $\lim_{n\to\infty}\lim_{N\to\infty}\mathbf{P}_{N,n}=\mathbf{P}_{S,0,x+i}^{chordal}$, integrated against $\rho(x)$. 
Given $\xi\geq 0$, by equation (\ref{eq:rightmost_conjecture}), we should have (approximately) 
\begin{equation}
\mathbf{P}_{N,n}(X<\xi)=\int_{-\infty}^\xi\left[1-\left(\frac{a(x,\xi)}{c(x,\xi)}\right)^2\right]^{5/8}\rho(x)\dx. \label{eq:final}
\end{equation}

We use numerical integration to calculate the right hand side of (\ref{eq:final}). Figure \ref{fig:rightmost_cdf}
shows a plot of the cumulative distribution function for $X$ under the measure $\mathbf{P}_{N,n}$ obtained from our
simulations, along with the conjectured cumulative distribution function for $X$ given by the right hand side of
(\ref{eq:final}), for values of $\xi$ between $0$ and $5$. In the scale of the figure, the two curves look almost
identical. In Figure \ref{fig:rightmost_difference}, we plot the difference between the simulated cdf for $X$ and
the conjectured cdf for $X$.

\begin{figure}[!htb]
\begin{center}
\includegraphics[height=10cm]{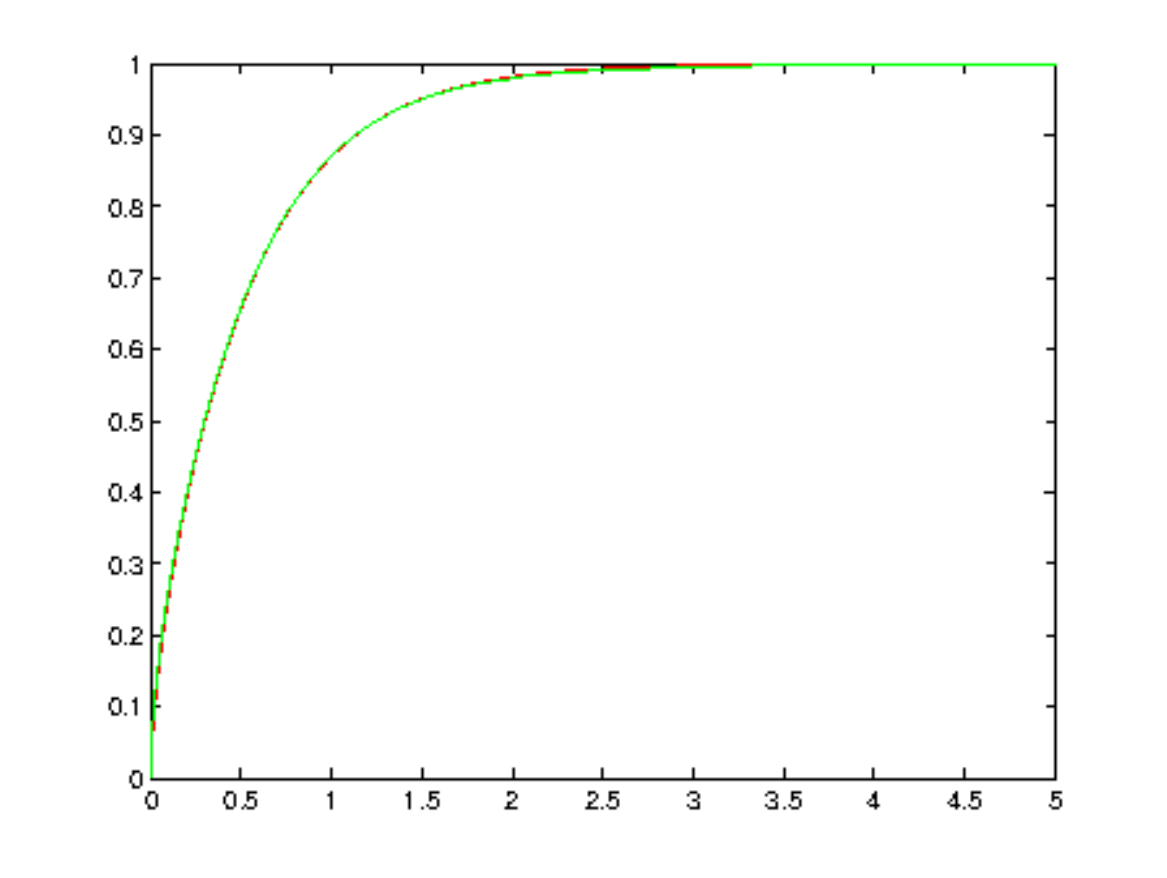}
\caption{Plot of the conjectured cdf for the rightmost excursion of SAW in the strip in the scaling limit as $\delta\to 0+$
and the simulated rightmost excursion for SAW in the fixed irreducible bridge ensemble. The conjectured cdf is colored
in red, while the simulated cdf is colored in green. In the scale of the image, it is difficult to see the difference.\label{fig:rightmost_cdf}}
\end{center}
\end{figure}

\begin{figure}[!htb]
\begin{center}
\includegraphics[height=10cm]{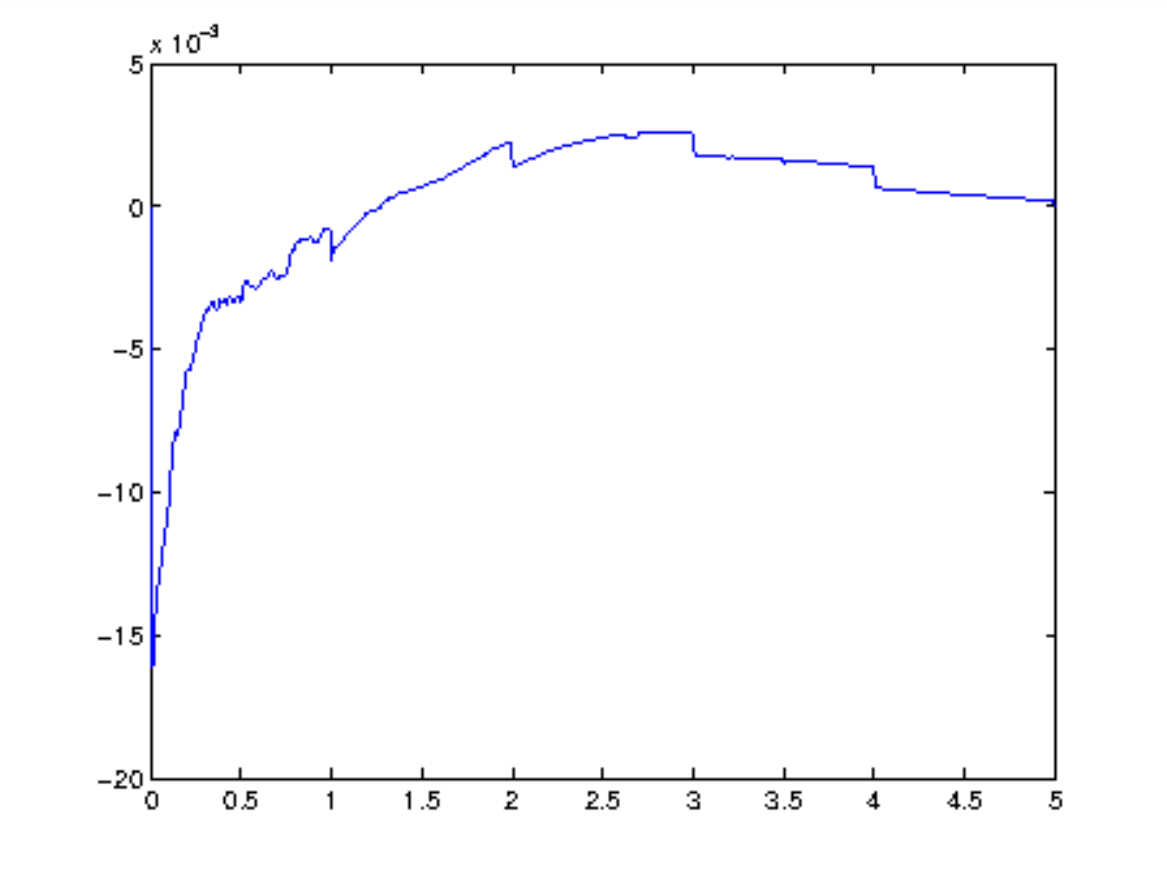}
\caption{Plot of the difference in values for the conjectured cdf for the rightmost excursion and the simulated cdf for
the rightmost excursion. We subtracted the simulated values for the cdf from the conjectured values. The error is small, but one 
can see that the error is larger near $x=0$. This is because there is a systematic error present due to the fact that we 
have chosen a finite value for $n$. There is a slight bias for bridge points falling closer to $x=0$ with a finite $n$. Choosing 
values of $n$ much higher than $n=100$ also creates some issues, since the number of one-million step SAWs with the given number 
of bridge points is drastically reduced for larger values of $n$. \label{fig:rightmost_difference}}
\end{center}
\end{figure}

%% file: appx1.tex
\newpage
\section{Appendix: Argument that $\sigma=4/3$} \label{appx:sigma}

Here we give an argument, originally due to Tom Kennedy via private communication, that the value of $\sigma$ defined by (\ref{eq:sigma}) is 4/3. Let $B_h(z)$ denote the generating function for bridges starting at the 0 with $h$. That is, if $\omega\in\mathcal{B}_n$, we set $h(\omega)=\textrm{Im}(\omega(n))$ and 
\begin{equation*}
B_h(z) = \sum_{\omega\in\mcB}z^{|\omega|}1(h(\omega)=h). 
\end{equation*}  

In \cite{lawler2002scaling}, Lawler Schramm and Werner conjecture that we should have $B_h(z)\asymp h^{-1/4}$, where here $\asymp$ means that the ratio of both sides are bounded away from $0$ and $\infty$. The argument goes as follows: If we constrain a bridge to have an endpoint at a fixed point of height $h$, the decay goes like $h^{-2b}$, where $b=5/8$. The number of endpoints that contribute to $B_h(z)$ is of order $h$. 
We begin by running an i.i.d. sequence of irreducible bridges until the concatenation of which has height which strictly exceeds some number $L>0$. This happens with probability $1$, and therefore we have 

\begin{align*}
1 & = \sum_{n=1}^\infty \sum_{\omega^1,\ldots,\omega^n\in\mcI}\mu^{-\sum_{j=1}^n|\omega^j|}1\left(\sum_{j=1}^n h(\omega^j)>L\right)1\left(\sum_{j=1}^{n-1}h(\omega^j)\leq L\right) \\
  & = \sum_{h=0}^L\sum_{n=1}^\infty\sum_{\omega^1,\ldots,\omega^n\in\mcI} \mu^{-\sum_{j=1}^n|\omega^j|}1\left(\sum_{j=1}^{n-1}h(\omega^j)=h\right)1\left(\sum_{j=1}^nh(\omega^j)>L\right) \\ 
  & = \sum_{h=1}^L\sum_{n=1}^\infty\sum_{\omega^1,\ldots,\omega^{n-1}\in\mcI}\mu^{-\sum_{j=1}^{n-1}|\omega^j|}1\left(\sum_{j=1}^{n-1}h(\omega^j)=h\right)\sum_{\omega^n\in\mcI}\mu^{-|\omega^n|}1\left(h(\omega^n)+h>L\right) \\
  & = \sum_{h=0}^LB_h(\mu^{-1})\sum_{\omega\in\mcI}\mu^{-|\omega|}1(h(\omega)>L-h). 
\end{align*}

Now we use 

\begin{equation} 
\sum_{\omega\in\mcI}\mu^{-|\omega|}1(h(\omega)>L-h) = \mathbf{P}(h(\omega)>L-h), \label{eq:appx-prob}
\end{equation} where we are using $\mathbf{P}$ to denote the probability measure on $\mcI$ defined by $\mathbf{P}(\omega)=\mu^{-|\omega|}$. We would like to develop a relationship between $B_h(z)$, and the cumulative distribution function for the height of an irreducible bridge. Using (\ref{eq:appx-prob}), we have 

\begin{equation} 
1 = \sum_{h=0}^LB_h(\mu^{-1})\mathbf{P}(h(\omega)>L-h). 
\end{equation} Let us now assume that $B_h(\mu^{-1})\asymp h^{-1/4}$ and $\mathbf{P}(h(\omega)>h)\asymp h^{-p}$ for some power $p$. We will split the above identity into two sums: one from $0$ to $L/2-1$, and one from $L/2$ to $L$. In the first sum, $L-h/2$ is at least $L/2$, and so $\mathbf{P}(h(\omega)>L-h)$ is (up to multiplicative constants) $L^{-p}$. So the first sum behaves like 
\begin{equation*}
\sum_{h=0}^{L/2-1}h^{-1/4}L^{-p}\asymp L^{-p+3/4}. 
\end{equation*} In the second sum, $h\geq L/2$ and so $B_h(\mu^{-1})$ is (up to multiplicative constants) $L^{-1/4}$. So the second sum behaves like 
\begin{equation*}
\sum_{L/2}^L L^{-1/4}(L-h)^{-p} = \sum_0^{L/2}L^{-1/4}h^{-p} \asymp L^{3/4-p}, 
\end{equation*} so both sums behave like $L^{3/4-p}$. As $L\to\infty$, the identity says that this cannot diverge or go to zero, and so we should have $p=3/4$. 

In conclusion, $\mathbf{P}(h(\omega)>h)$ decays like 
\begin{equation*}
\mathbf{P}(h(\omega)>h)\asymp h^{-3/4}. 
\end{equation*} This tells us which stable process the sum of $n$ irreducible bridges converges to in distribution. Let $Y_n$ denote the $n$-th bridge height. We want to find $\sigma$ so that $Y_n$ grows like $n^\sigma$. The cdf $F(h)$ of the irreducible bridge heights converges to 1 like $1-h^{-3/4}$ as $h\to\infty$. If there are $n$ irreducible bridges, the larges one will roughly have height $h$, so $F(h)\approx 1-1/n$. Thus $h\sim n^{4/3}$, i.e. $\sigma=4/3$. 

